# lightcurver: A Python Pipeline for Precise Photometry of Multiple-Epoch Wide-Field Images

Frédéric Dux [1,2]

**1** European Southern Observatory, Alonso de Córdova 3107, Vitacura, Santiago, Chile **2** Ecole Polytechnique Fédérale de Lausanne (EPFL), Switzerland

## Summary

`lightcurver` is a photometry pipeline for time series astronomical imaging data, designed for the semi-automatic extraction of precise light curves from small, blended targets. Such targets include, but are not limited to, lensed quasars, supernovae, or Cepheids in crowded fields. `lightcurver` is not a general purpose photometry, astrometry, and classification pipeline like `legacypipe` (Legacy Surveys Collaborations, 2024). Instead, it is a framework tailored to the precise study of a small region of interest (ROI) in wide-field images, utilizing stars surrounding the ROI to calibrate the frames.

At its core, `lightcurver` leverages STARRED (Michalewicz et al., 2023; Millon et al., 2024) to generate state-of-the-art empirical point spread function (PSF) models for each image. It then determines the relative zeropoints between images by combining the PSF-photometry fluxes of several stars in the field of view. Subsequently, STARRED is used again to simultaneously model the calibrated pixels of the ROI across all epochs. This process yields light curves of the point sources and a high-resolution image model of the ROI, cumulating the signal from all epochs.

`lightcurver` aims to be maintainable, fast, and incremental in its processing approach. As such, it can enable the daily photometric analysis of a large number of blended targets in the context of the upcoming Rubin Observatory Legacy Survey of Space and Time (LSST; Vera C. Rubin Observatory LSST Solar System Science Collaboration et al., 2021).

## Statement of need

The LSST survey will generate an unprecedented amount of imaging data, revisiting the same regions of the sky every four days, with irregular pointings due to its observing strategy. Processing data at this cadence will require robust pipelines capable of ingesting new observations and providing immediate photometric calibration and analysis. This is particularly important for time-sensitive targets of opportunity, where rapid reaction to changes is essential for timely follow-up. An existing pipeline that performs this precise deblending and photometric measurement task, COSMOULINE (Magain et al., 1998; The COSMOGRAIL collaboration, 2010), requires too much manual intervention to be run on a daily basis.

On the other hand, STARRED is a powerful PSF modelling and deconvolution package, ideal for this task. However, by its nature, it cannot include an infrastructure that makes it convenient to apply to large datasets without manual intervention (e.g., visually identifying appropriate stars, extracting cutouts, and all subsequent processing steps leading to a light curve). Particularly, STARRED modelling requires a very stable zeropoint across modelled epochs, as it emulates the constant components of the ROI as one grid of pixels common to all epochs, which it simultaneously optimizes together with the fluxes of the variables. Achieving such precise relative zeropoint calibration (typically one millimag), especially in an automated manner, comes with challenges.

`lightcurver` addresses this challenge by automatically selecting calibration stars, modelling them, and robustly combining their fluxes to calibrate the zeropoints, making it suitable as a daily running pipeline on a large number of ROIs.

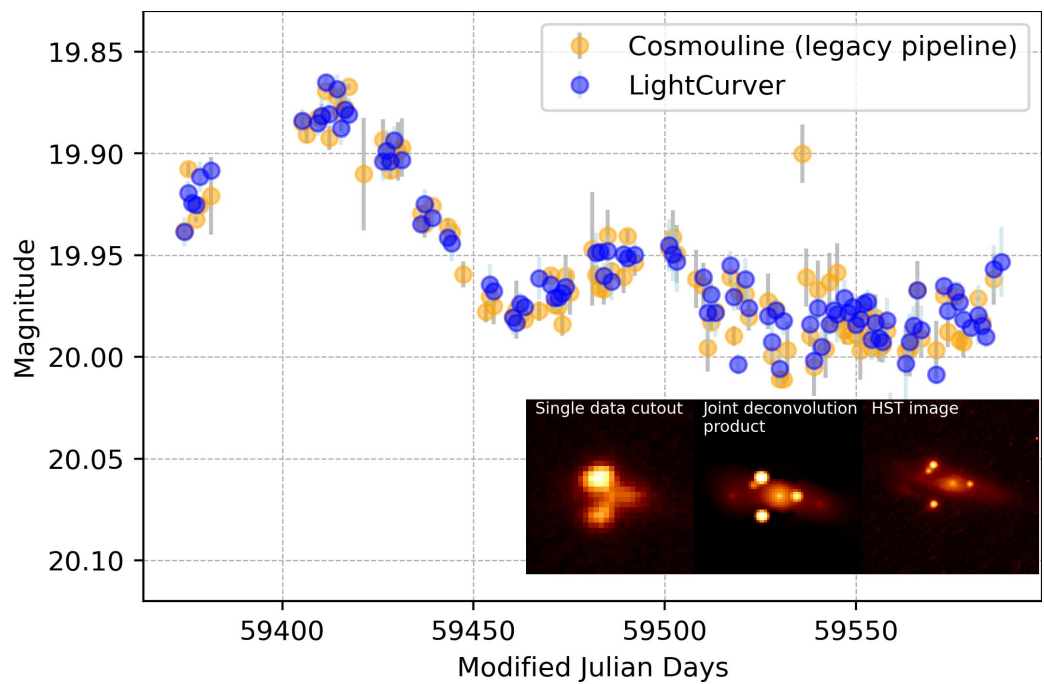

**Figure 1:** Light curve of a lensed image of a quasar (J0030-1525), extracted once with the existing code base (`COSMOULINE`), requiring a week of investigor's time, and another time with `lightcurver`, requiring about an hour of investigator's time. HST image: PI Tommaso Treu, proposal GO 15652 (Treu, 2018).

## Functionality

`lightcurver` utilizes an SQLite3 database to track data processing stages and relies on SQL queries to manage its workflow, identifying the processing required at each step. First, the frames undergo background subtraction, and the sources are extracted using `sep` (Barbary, 2016; Bertin & Arnouts, 1996). The positions of the extracted sources are then used to plate-solve each frame, primarily with `Astrometry.net` (Lang et al., 2010). This permits sanity checks with `pyephem` (Rhodes, 2011), but also allows for an automatic selection of calibration stars around the ROI by querying Gaia (Gaia Collaboration, 2016) with `astroquery` (Ginsburg et al., 2019) for suitable stars. The pointings and field rotations do not need to be stable across epochs, as each frame is assigned its own calibration stars with the help of `shapely` (Gillies et al., 2024).

Subsequently, cutouts of the ROI and stars are extracted using `astropy` (Astropy Collaboration, 2013; Astropy Collaboration et al., 2018, 2022), masked, cleaned from cosmic rays with the help of `astroscrappy` (McCully et al., 2018; van Dokkum, 2001), and stored in an HDF5 file (Fortner, 1998). The PSF model is then calculated for each frame with `STARRED` before being stored in the same HDF5 file. Next, the fluxes of all the calibration stars in all frames are measured using PSF photometry, and the resulting fluxes of the stars are scaled and combined to obtain precise relative zeropoints. To avoid clearly failed fits from corrupting the reduction, each fitting procedure (PSF or photometry) stores its reduced chi-squared statistic in the database, allowing downstream steps to filter which frame, PSF, or flux to proceed with.

The software is divided into three main sub-packages: `processes`, containing individual data processing tasks; `pipeline`, defining the sequence of these tasks to ensure orderly data analysis;

and `structure`, containing the database schema and handles user configuration. Users can customize the processing of datasets through a YAML configuration file, allowing flexibility in handling various data characteristics. Typically, the YAML configuration file needs to be configured once when executing the pipeline on the first few frames, but the subsequent addition of new frames as they are observed requires no further manual intervention.

`lightcurver`, in comparison to `COSMOULINE`, achieves equal or better photometric precision in a much more automated fashion. Figure 1 presents the light curve of the southernmost lensed image of the quasar J0030-1525 ([Lemon et al., 2018](#)), extracted from the same dataset (ESO program 0106.A-9005(A), PI Courbin) using both `COSMOULINE` and `lightcurver`. The stable zeropoint across frames enables STARRED to reliably fit the constant components, in this case, two galaxies visible in the image. This reliable deblending of the different flux components yields both light curves and a high resolution image, the morphology of which is confirmed by comparison with Hubble Space Telescope imaging.

In summary, `lightcurver` is a robust and efficient photometry pipeline designed for the semi-automatic extraction of precise light curves from small, blended targets in cadenced astronomical imaging data. By leveraging the power of STARRED for state-of-the-art PSF modelling and deconvolution, and employing an automated flux calibration process, `lightcurver` achieves equal or better photometric precision compared to existing pipelines, while requiring significantly less manual intervention.

## Acknowledgments

Special thanks to Martin Millon, Malte Tewes, Vivien Bonvin and Frederic Courbin for conceptualizing and creating the `COSMOULINE` pipeline. While this code base it not a re-write of `COSMOULINE`, it certainly shares the philosophy of some of the processing steps.

I am grateful to the two reviewers, Tom Wilson and Benjamin Rose, for sharing their expertise in identifying and reporting potential issues in `lightcurver`, and to the JOSS team for making this publication platform possible.

This project also made use of some of the backbone packages of scientific Python computing: NumPy ([Harris et al., 2020](#)), SciPy ([Virtanen et al., 2020](#)), Matplotlib ([Hunter, 2007](#)) and Pandas ([McKinney, 2010](#); [pandas, 2020](#)). For first guess photometry, this software also benefited from the functions of `photutils` ([Larry Bradley et al., 2024](#)), and for finding transformations between frames, `astroalign` ([Beroiz et al., 2020](#)).